\documentclass[twocolumn]{article}
\usepackage[utf8]{inputenc}
\usepackage{amssymb, amsmath}
\usepackage{color}
\usepackage[backend=biber,sorting=ynt]{biblatex}
\usepackage[
  a4paper,
  left=0.6in,
  right=0.6in,
  top=0.75in,
  bottom=0.75in
]{geometry}
\usepackage{array}  % For the 'm' column type
\usepackage{setspace}
\usepackage[table]{xcolor}
\usepackage{graphicx}
\usepackage{subcaption}
\usepackage{multirow}
\usepackage{algorithm}
\usepackage{algpseudocode}
\usepackage{enumerate}

\algrenewcommand\algorithmicthen{}

\usepackage{amsthm}
\newtheorem{definition}{Definition}

\title{Beyond Lemma Sharing -- Novel Parallelization Strategies for Property Directed Reachability}
\author{Verner Vla\v{c}i\'{c} \\
\small Huawei Technologies Switzerland AG \\
\small verner.vlacic1@huawei.com}
\date{Feb 2026}

\begin{document}

\maketitle

\begin{abstract}
Property Directed Reachability (PDR) is a commonly used technique for automated hardware model checking, yet efficiently parallelizing it remains a significant challenge. Existing approaches, such as lemma sharing, often suffer from limited scalability as processor counts increase. In this work, we present two novel sharing-based parallelization strategies, \emph{preemptive propagation} and ARPOS, and compare their performance with classical lemma sharing.

To this end, we develop an asynchronous MPI-based message passing framework for the state-of-the-art rIC3 hardware model checker. Experimental results on the 2025 Hardware Model Checking Competition benchmark demonstrate that our \emph{preemptive propagation} strategy yields a significant performance boost over classical lemma sharing.
\end{abstract}

\section{Introduction}

First introduced as IC3 in \cite{Bradley2011SATBasedMC}, Property Directed Reachability (PDR) is a sophisticated algorithm for automated model checking that incrementally constructs an inductive proof of unreachability of fault states. It is a core component in most modern hardware safety checking tools, as reflected by its central role in the Hardware Model Checking Competition \cite{hwmcc}. Beyond hardware, PDR has also been successfully adapted to software verification \cite{pdr-software-2020} and to related domains such as automated planning and scheduling \cite{pdr-auto-planning-2014}.

PDR relies on accumulating reachability information by constructing and solving ancillary 1-step transition problems.
In most applications (automated planning \cite{pdr-auto-planning-2014} constituting a notable exception), these ancillary problems are SAT queries. While the general Boolean satisfiability problem is NP-complete, the theoretical and practical importance of automated model checking have nevertheless motivated sustained research and engineering efforts aimed at improving the performance of PDR \cite{EenPDR2011}. These efforts have enabled PDR-based tools to routinely solve verification problems that were previously out of reach, even as the size and complexity of industrial benchmarks continue to grow.

A number of significant refinements to the original PDR algorithm have been proposed over the past decade. Among the most influential are counterexamples to generalization (CTG) \cite{CTG2013}, which improves clause generalization, the use of internal signals \cite{InternalSignals2021}, localization abstraction \cite{LocalizationAbs2017} to reduce the effective state space, and the introduction of the inductive frame \cite{PushingToTheTop2015}. In the hardware model checking domain, the current state-of-the-art implementation of PDR is rIC3, champion of the 2025 edition of the Hardware Model Checking Competition \cite{hwmcc}. Introduced in \cite{rIC3-2025}, rIC3 is developed in Rust and relies on a specialized SAT solver, GipSAT, which is deeply optimized for the characteristic structure of SAT queries in PDR.

Despite these advances, efficient use of parallelism remains a major challenge in further scaling automated model checking. In particular, exploiting parallelism within PDR is difficult due to its highly complex control-flow logic and very limited opportunities for data parallelism. As a result, there has been comparatively little research on parallelizing PDR \cite{chakikarimi2016, marescotti2017, clifton2022fast}.

In this work, we revisit existing parallelization strategies for PDR---most notably \emph{Lemma Sharing}---and introduce \emph{preemptive propagation}, an optimization designed to reduce redundant SAT queries across parallel engines. We further propose a novel sharing-based parallelization strategy, \emph{asynchronous rescheduled proof obligation sharing} (ARPOS). To support these techniques, we develop a Rust-based framework for asynchronous point-to-point messaging over MPI, enabling their integration with the state-of-the-art rIC3 hardware model checking engine. We evaluate our approach on the bit-level safety track of the 2025 Hardware Model Checking Benchmark \cite{hwmcc}. Our results show that \emph{preemptive propagation} significantly improves the performance of both classical Lemma Sharing as well as ARPOS at higher processor counts.

The remainder of this paper is organized as follows. Section \ref{sec:related-work} reviews related work. Sections \ref{sec:pdr-review} and \ref{sec:port-and-lemshare} provide a brief overview of the sequential PDR algorithm and standard parallelization strategies, respectively. Section \ref{sec:preempt-prop}
introduces our preemptive propagation optimization, and Section \ref{sec:ARPOS} presents the design of ARPOS and compares it with existing approaches. Implementation details of our framework are described in Section \ref{sec:implementation}, followed by the experimental evaluation in Section \ref{sec:experiments}. Finally, Section \ref{sec:conclusion} concludes the paper.

\section{Related Work}\label{sec:related-work}

The simplest and most commonly employed parallelization strategy for PDR is the so-called \emph{portfolio}. In this setting, several safety-checking engines are run in an embarrassingly parallel manner, each corresponding to a different configuration of a sequential PDR implementation, although other safety-checking algorithms such as k-induction \cite{kinduction2000} can also be included. The portfolio terminates as soon as any engine produces either a \emph{Safe} or \emph{Unsafe} result, making it easy to implement and often effective in practice as it combats the large observed variance of the runtime when running sequential PDR on the same problem with different configurations.

Beyond Portfolio, several authors have investigated \emph{lemma sharing} as a more tightly coupled form of parallelization. This idea was already hinted at in the original IC3 paper \cite{Bradley2011SATBasedMC} and was later further explored in \cite{chakikarimi2016, marescotti2017}. In these approaches, parallel workers exchange inductive clauses or lemmas in order to preclude redundant work performed by multiple engines. 
Our proposed method ARPOS builds on this line of research. For completeness, we review the basic lemma-sharing strategy in Section \ref{sec:lemma-sharing}.

Marescotti et al. \cite{kinduction2000} further explored combinations of Lemma Sharing with \emph{property partitioning}. In this strategy, the safety property is effectively
``unrolled'' by applying the transition relation and then decomposed syntactically into a set of sub-properties. The original verification problem is safe if and only if each of these subproblems is safe, allowing different workers to reason about different portions of the property space while still benefiting from shared information. This work is implemented using the Spacer model checker \cite{komuravelli2016anvesh} and evaluated on a standard software verification benchmark.

More recently, Clifton and Gretton proposed PD-PDR \cite{clifton2022fast}, a parallel architecture in which a central coordinator maintains the global state of the PDR computation and distributes individual SAT queries to worker processes. While this approach demonstrates the potential of fine-grained parallelism, it was primarily developed and evaluated in the context of automated planning rather than hardware or software model checking.

We discuss how our new strategy ARPOS relates to these works in Section \ref{sec:comp-w-other-methods}.

\section{Property Directed Reachability}\label{sec:pdr-review}
In this paper, we study the verification of safety properties for finite-state transition systems, which we now define. For a set of Boolean variables $X$ describing a part of the system state, we write $X'=\{x' \vert x\in X\}$ to denote the corresponding variables at the next time step. Similarly, for a Boolean formula $\phi$, we use $\phi'$ to denote the formula obtained by replacing each variable in $\phi$ with its primed counterpart.

A \emph{literal} over a Boolean variable $x$ is either $x$ or its negation $\neg x$. A disjunction of literals is called a \emph{clause}, and a Boolean formula is said to be in \emph{conjunctive normal form} (CNF) if it is a conjunction of clauses.

\subsection{Safety properties}

A transition system over a set of Boolean-valued variables $X$ is described by a Boolean formula $Init(X)$, representing the set of initial states, and a Boolean formula $T(X,X')$, encoding the possible transitions from the current-state variable valuations to the next-state valuations. A sequence of states $(s_1,s_2,\dots, s_n)$, where each $s_i$ is a valuation of the variables in $X$, is called a \emph{path} of the transition system if $(s_{i},s_{i+1}) \models T$, for all $i$.

Given another Boolean formula $P(X)$, called the \emph{safety property}, we say that the transition system is \emph{$P$-invariant} if every state reachable from an initial state along some path satisfies $P$.

\subsection{Frames}
We now give a high-level overview of the PDR algorithm, originally introduced in \cite{Bradley2011SATBasedMC} as IC3 and subsequently refined and extended in various works \cite{EenPDR2011, PushingToTheTop2015, CTG2013, rIC3-2025}.

\begin{definition}[Frames]
Given a transition system $(Init(X), T(X,X'))$ and a safety property $P(X)$, a PDR trace is a sequence $(F_0, F_1, \dots, F_n)$, where each $F_i$ is a Boolean formula in CNF such that
\begin{align}
    F_0 &\equiv Init \label{eq:F1}\\
    F_i \wedge T &\implies F_{i+1}' \label{eq:F2}\\
    F_i &\implies F_{i+1} \label{eq:F3}\\
    i < n &\implies (F_{i} \implies P) \label{eq:F4}
\end{align}
\end{definition}

Each formula $F_i$ is called a \emph{frame} and corresponds to an over-approximation of the set of states reachable from the initial states in at most $i$ transitions. The clauses comprising the frames are commonly referred to as \emph{lemmas}.

The PDR algorithm begins with the single frame $(F_0)$ and incrementally constructs additional frames with the goal of reaching a fixed point $F_{n-1}= F_n$, which, due to the conditions \eqref{eq:F1}--\eqref{eq:F4}, implies that $P$ is an invariant of the transition system. Such a frame is called an \emph{inductive strengthening} of $P$. 
If no such strengthening can be found, the algorithm instead produces a path from $Init$ to $\neg P$, called a \emph{counterexample trace}, demonstrating that $P$ is not invariant.

Concretely, in each top-level iteration $n$, PDR attempts to strengthen the frames by adding lemmas so that the current top-level frame $F_n$ becomes safe, i.e., $F_n \implies P$. This condition is tested by issuing the SAT query
\begin{equation}
     \mathsf{SAT}\left( F_{n} \wedge \neg P \right) \label{eq:top-level}
\end{equation}
If this query is satisfiable, the corresponding state needs to be \emph{blocked} in the \emph{blocking phase}, described next.
If, on the other hand, \eqref{eq:top-level} is unsatisfiable, a new frame $F_{n+1} \equiv \top$ is introduced which initially contains no clauses, and the current top-level iteration is concluded with the \emph{propagation phase}, in which  existing lemmas are tested for relative inductiveness and propagated forward whenever possible.

\subsection{Proof obligations}

\begin{definition}[Proof Obligation] Let a transition system $(Init(X), T(X,X'))$, a safety property $P(X)$, and a PDR trace $(F_0, F_1, \dots, F_n)$ be given. A \emph{proof obligation} is a pair $\langle s, \ell\rangle$ where $s$ is a cube over state variables and $\ell \leq n$ is an integer called the \emph{level}, such that:
\begin{enumerate}
    \item $s \wedge F_\ell$ is satisfiable, and
\item
every state $m \models s$ can reach a state violating $P$ along some path of the transition system.
\end{enumerate}
\end{definition}
The presence of a proof obligation $\langle s, \ell\rangle$ indicates that the states described by $s$ must be blocked from all frames $F_i$ with $i\geq \ell$ in order for the trace to have a chance of converging to an inductive strengthening of $P$. In particular, if a proof obligation $\langle s, 0 \rangle$ is encountered such that $\mathsf{SAT}(s \wedge Init)$ holds, then there exists a path from an initial to a bad state, and the algorithm terminates with the result \textit{Unsafe}.

To handle a proof obligation
$\langle s,\ell\rangle$ with $\ell \geq 1$, PDR issues the SAT query
\begin{equation}
    \mathsf{SAT}\left( F_{\ell-1} \wedge \neg s \wedge T \wedge s' \right). \label{eq:inductivity}
\end{equation}
If the query is unsatisfiable, $s$ undergoes a \emph{generalisation} procedure---typically by dropping literals---to obtain a bigger cube $c$, and the clause $\neg c$ is added to the frames $F_1, F_2, \dots, F_\ell$. This excludes $s$ from $F_\ell$ while preserving the conditions \eqref{eq:F1}--\eqref{eq:F4}. If, on the other hand, \eqref{eq:inductivity} is satisfiable with model $m$, then $m$ must first be blocked from $F_{\ell-1}$ before attempting to block $s$ from $F_\ell$ again. Accordingly, a new proof obligation $\langle m, \ell - 1\rangle$ is created and enqueued.

Hence, in addition to the trace, the PDR algorithm maintains a priority queue $\mathcal{Q}$ of proof obligations, where obligations at lower levels are assigned higher priority. In order to streamline the exposition of our parallel strategy ARPOS for PDR in Section \ref{sec:ARPOS}, we decompose the blocking phase into two procedures. The procedure $\textsc{HandleOne}$ processes a single proof obligation, while $\textsc{Block}$ handles all proof obligations in the queue $\mathcal{Q}$ by repeatedly invoking $\textsc{HandleOne}$. These procedures are specified in pseudocode in Algorithm \ref{alg:handle-one} and Algorithm \ref{alg:block}, respectively.

The procedure $\textsc{HandleOne}$ can return one of three distinct return value types, corresponding to the three possible different outcomes of attempting to block a proof obligation.

\begin{algorithm} 
\caption{\textsc{HandleOne}}\label{alg:handle-one}
\begin{algorithmic}[1]
\Require trace $(F_0, F_1,\dots, F_n)$, proof obligation $\langle s,\ell\rangle$ with $\ell \leq n$
\If{$\ell=0$ and $\mathsf{SAT}(s \wedge Init)$}
    \State \textbf{return} Unsafe
\EndIf
% \If{$s$ already blocked by some $F_i$, $i\geq \ell$}
%     \State \textbf{return} Pushed$(i)$
% \EndIf
\If{$m \gets \mathsf{SAT}(F_{\ell-1} \wedge \neg s \wedge T \wedge s')$}
    \State \textbf{return} NewPO$(m)$
\Else
    \State $(c, k) \gets \Call{Generalize}{s}$
    \Comment{$s\Rightarrow c,\;  \ell \leq k$}
    \State \textbf{return} Generalized$(c, k)$
\EndIf
\end{algorithmic}
\end{algorithm}

\begin{algorithm}
\caption{\textsc{Block}}\label{alg:block}
\begin{algorithmic}[1]  % [1] = line numbers
\Require trace $(F_0, F_1,\dots, F_n)$, queue of proof obligations $\mathcal{Q}$ at level $n$ or above.
\While{$\langle s, \ell\rangle:=\Call{Peek}{\mathcal{Q}}$ and $\ell \leq n$}
    \State$\Call{Pop}{\mathcal{Q}}$
    \State $res \gets \Call{HandleOne}{(s,\ell)}$
    \If{$res=$ Unsafe}
        \State\textbf{return} Unsafe
    %\ElsIf{$res=$ Pushed($i$)}
    %\If{$res=$ Pushed($i$)}
    %     \State$\Call{Add}{\mathcal{Q}, \langle s, i+1\rangle}$
    \ElsIf{$res=$ NewPO$(m)$}
        % \If{$\ell=1$ and $\mathsf{SAT}(m \wedge Init)$}
        %     \State\textbf{return} False
        % \EndIf
        \State$\Call{Add}{\mathcal{Q}, \langle s, \ell\rangle}$
        \State$\Call{Add}{\mathcal{Q}, \langle m, \ell-1\rangle}$\label{ln:block-child-po}
    \ElsIf{$res=$ Generalized$(c,k)$ \label{ln:block-generalized}}
        \For{$i=1$ to $k$} \label{ln:block-lemma-added}
            \State $F_i \gets F_i \cup \{ \neg c \}$
        \EndFor
        \State $\Call{Add}{\mathcal{Q}, \langle s, k+1\rangle}$
       \label{ln:block-reschedule} \Statex\Comment{Reschedule $s$ to a higher frame}
    \EndIf
\EndWhile
\State\textbf{return} Success
\end{algorithmic}
\end{algorithm}

From the procedure \textsc{Block}, we observe that a proof obligation is rescheduled on $\mathcal{Q}$ every time it is successfully blocked, and each time at a higher level than before. This 
reflects the fact that any state that can reach $\neg P$ must ultimately be excluded from all frames. Because each new proof obligation is initially created at lower levels than the already existing ones, rescheduling causes multiple proof obligations to coexist at the same level. As we will show in Section \ref{sec:ARPOS}, this structural property can be exploited to introduce parallelism.
Moreover, once created, a proof obligation is eventually rescheduled at level $n+1$, and therefore persists as an element of $\mathcal{Q}$ into the next top-level iteration of PDR.
%When \textsc{Block} terminates successfully, the proof obligations remaining in $\mathcal{Q}$ are all at level $n+1$ and will be blocked in the next PDR top-level iteration.

\subsection{Propagation}
The propagation phase is conceptually simpler than the blocking phase. After a new frame $F_{n+1}$, initially empty, has been appended to the trace, the algorithm considers each clause $\neg c \in F_\ell$, for $\ell=1,\dots, n$, in increasing order of $\ell$, and issues the SAT query
\begin{equation}
    \mathsf{SAT}\left( F_{\ell} \wedge \neg c \wedge T \wedge c' \right). \label{eq:prop-query}
\end{equation}
If this query is unsatisfiable, the clause $\neg c$ 
is inductive relative to $F_\ell$ and can be propagated the next frame $F_{\ell+1}$ without violating conditions \eqref{eq:F1} -- \eqref{eq:F4}. If at any point during propagation two consecutive frames conincide, i.e., $F_{\ell}= F_{\ell+1}$ for some $\ell$, the algorithm has found the inductive strengthening of $P$ and returns \emph{Safe}. 

\begin{algorithm} 
\caption{\textsc{Propagate}}\label{alg:propagate}
\begin{algorithmic}[1]
\Require trace $(F_1,\dots, F_{n+1})$ with $F_{n+1} = \varnothing$
\For{$\ell=0$ to $n$}
    \For{each clause $\neg c$ in $F_\ell$}
        \If{$\neg \mathsf{SAT}(F_{\ell} \wedge \neg c \wedge T \wedge c')$}\label{ln:propagate-sat-query}
            \State $F_{\ell+1} \gets F_{\ell+1} \cup \{\neg c\}$\label{ln:propagate-lemma-added}
        \EndIf
    \EndFor
    \If{$F_{\ell} = F_{\ell + 1}$}\label{ln:propagate-safe}
        \State\textbf{return} Safe
    \EndIf
\EndFor
\end{algorithmic}
\end{algorithm}

\section{Portfolio and Lemma Sharing}\label{sec:port-and-lemshare}

We briefly review two parallelization strategies for PDR that have been explored in the literature for distributed scenarios \cite{marescotti2017}.

\subsection{Portfolio}
The simplest parallelization strategy is the \emph{portfolio} approach, in which multiple sequential PDR instances are executed independently on the same transition system and safety property. Each instance is capable of solving the verification problem on its own, and the overall execution terminates as soon as any instance returns either \emph{Safe} or \emph{Unsafe}.
The individual engines are run with different configurations, such as distinct SAT-solver random seeds or different combinations of enhancements, including CTG or localization abstraction \cite{CTG2013, LocalizationAbs2017}.

Despite its simplicity, the portfolio strategy is often highly effective, as it exploits the significant variability in the runtime of sequential PDR across different configurations on the same problem. Indeed, portfolio-based approaches are widely used in the Hardware Model Checking Competition \cite{hwmcc}, including the winning entry of the 2025 edition, rIC3 \cite{rIC3-2025}.

This variability arises from the inherent non-uniqueness of both solutions and search trajectories---both in the \emph{Safe} case of non-unique inductive strengthenings of $P$ and in the \emph{Unsafe} case of non-unique counterexample traces.

\subsection{Lemma sharing}\label{sec:lemma-sharing}
A more tightly coupled parallelization strategy studied in the literature is \emph{Lemma Sharing} \cite{Bradley2011SATBasedMC, chakikarimi2016, marescotti2017}. As in the portfolio approach, multiple PDR engines run in parallel, each maintaining its own trace and proof obligation queue. However, whenever an engine derives a new lemma (line \ref{ln:block-generalized} in Algorithm \ref{alg:block}), that lemma is communicated to the other engines and incorporated into their PDR traces.

The motivation behind Lemma Sharing is to reduce redundant work: different engines may encounter closely related proof obligations, and sharing lemmas allows them to benefit from each other's progress. As a result, although the engines operate independently, their traces are kept closely aligned throughout the computation.

Our first algorithmic contribution builds on this setting by introducing an optimization of the propagation phase, which we describe next.

\section{Preemptive propagation}\label{sec:preempt-prop}
Our first optimization targets the propagation phase of PDR in the presence of lemma sharing. Suppose an engine receives a message $(c, \ell+1)$ indicating that the clause $\neg c$ has been propagated to frame $F_{\ell + 1}$ elsewhere. If this message arrives before the local engine considers $c$ in the propagation query \eqref{eq:prop-query}, then the query will be skipped, saving computation time on this engine.

However, in standard Lemma Sharing such a message is sent only when $c$ is successfully propagated i.e., when the query is \emph{unsatisfiable}. If, on the other hand, propagation fails (the query is satisfiable), then no message is sent and each engine must independently solve \eqref{eq:prop-query} and obtain the same satisfiable result. This redundancy is particularly costly because, in practice, satisfiable queries are both more frequent and more expensive than unsatisfiable ones in PDR \cite{EenPDR2011}.

To mitigate this inefficiency, we extend the message format. Whenever an engine solves $\eqref{eq:prop-query}$ for a clause $c$, it sends the message $(c, \ell+1, \emph{succ})$ to other engines, where the Boolean flag $\emph{succ}$ indicates whether the propagation of $c$ failed or succeeded. Thus, if this message is received on another engine before $c$ is considered for propagation locally, the corresponding SAT query is skipped and, if $\emph{succ}=\emph{true}$, the clause $\neg c$ is adjoined to the local $F_{\ell+1}$.
We refer to this optimization \emph{preemptive propagation}.

\medskip

Care must be taken, however, since preemptive propagation can inadvertently prevent useful propagations. Consider two engines, \emph{E1} and \emph{E2}, with respective frames $F_\ell^{1}$ and $F_{\ell}^2$ such that $F_\ell^{1}\neq F_\ell^{2}$. This situation may arise in a distributed setting, where traces are not guaranteed to remain identical at all times. Suppose further that $\neg c$ belongs to both frames, but $F_{\ell}^{1} \wedge \neg c \wedge T \wedge c'$ is satisfiable, whereas $F_{\ell}^{2} \wedge \neg c \wedge T \wedge c'$ is unsatisfiable. According to the preemptive propagation strategy, $\emph{E1}$ sends the message $(c, \ell+1, \emph{false})$. If $\emph{E2}$ receives this message before it gets the chance to issue its own corresponding SAT query for $c$, it skips propagation of $\neg c$, even though \emph{E2} would have successfully propagated it. By contrast, under classical Lemma Sharing, no message would have been sent by \emph{E1}, and \emph{E2} would correctly propagate $c$.

\begin{algorithm}[ht]
\caption{\textsc{PreemptPropagate}}\label{alg:preempt-propagate}
\begin{algorithmic}[1]
\Require trace $(F_1,\dots, F_{n+1})$ with $F_{n+1} = \varnothing$,\; message dictionary
$\textsc{Consult}:\mathrm{Clauses}\times\mathrm{Levels} \to \mathrm{Boolean}$,\;
priority function $\textsc{Prio}:\mathrm{Clauses} \to \mathrm{Integers}$
\State $\Call{SynchronizeEngines}{\,}$\label{ln:preempt-sync}
\For{$\ell=0$ to $n$}
    \State sort $F_\ell$ according to $\textsc{Prio}$
    \For{each clause $\neg c$ in $F_\ell$}
    \State $inbox\gets \Call{ReceiveMessages}{\,}$
        \For{$(\tilde{c},\tilde{\ell},\emph{succ})$ in $inbox$}
            \State$\Call{Consult}{\tilde{c},\tilde{\ell}} \gets \emph{succ}$
        \EndFor
        \If{$\Call{Consult}{c,\ell}=true$}
            \State $F_{\ell+1} \gets F_{\ell+1} \cup \{\neg c\}$
        \ElsIf{$\Call{Consult}{c,\ell}=false$}
            \State continue
        \ElsIf{$\neg \mathsf{SAT}(F_{\ell} \wedge \neg c \wedge T \wedge c')$}\label{ln:propagate-sat-query}
            \State $F_{\ell+1} \gets F_{\ell+1} \cup \{\neg c\}$
            \State$\Call{SendLemmaToAll}{(c,\ell + 1,\emph{true})}$
        \Else
            \State$\Call{SendLemmaToAll}{(c,\ell + 1,\emph{false})}$
        \EndIf
    \EndFor
    \If{$F_{\ell} = F_{\ell + 1}$}\label{ln:propagate-safe}
        \State\textbf{return} Safe
    \EndIf
\EndFor
\end{algorithmic}
\end{algorithm}

Failing to propagate lemmas does not compromise the soundness of the algorithm, but is detrimental to its performance, since missing a lemma in later stages of the PDR computation can make it difficult to fulfill the safety condition (line \ref{ln:propagate-safe} in Algorithm \ref{alg:propagate}), thus increasing the number of top-level iterations required to establish safety.

We remedy this issue by introducing a controller-worker architecture in which the controller acts as a buffer for all shared lemmas. The controller ensures that, if engine $E1$ requests to enter propagation phase $n$ before $E2$, then all lemmas from the trace of $E1$ are shared with $E2$ before $E2$ enters the same propagation phase. Consequently, after propagation, the trace of each engine contains at least the lemmas that would have resulted by propagating---without preemptive propagation---the lemmas of the first engine to enter the propagation phase.

Finally, to promote load balancing and reduce redundant work, engine visits the clauses in $F_\ell$ in a different order. Concretely, suppose there are $p$ engines with identifiers $0, 1, \dots, p-1$. We fix a deterministic hash function $h$ that maps clauses to integers. 
On engine $r$, the clauses $\neg c \in F_\ell$ are sorted according to the priority function 
\begin{equation}\label{eq:priority-function}
\textsc{Prio}_{r,p}(\neg c) = \left(h(\neg c) + r \right)\mod p
\end{equation}
This reduces the likelihood that different engines attempt to propagate the same clause simultaneously, since all engines traverse $F_\ell$ in the same cyclical order but start from positions that are spaced apart by $\#(F_\ell)/p$ on average.

Putting everything together, we obtain the method \textsc{PreemptPropagate} specified in pseudocode in Algorithm \ref{alg:preempt-propagate}.

\section{ARPOS -- Asynchronous Rescheduled Proof Obligation Sharing }\label{sec:ARPOS}

We propose a new parallelization strategy for PDR, called \emph{asynchronous rescheduled proof obligation sharing} (ARPOS). The key motivation for this method is the following observation.
In a standard Lemma Sharing scenario, each time an engine completes a call to \textsc{Block}, it identifies a new bad state, say $b$, at the top-level frame $F_n$ by solving the query \eqref{eq:top-level}. During this time, however, other engines are occupied with blocking proof obligations at lower levels of the trace.

Crucially, the blocking of these lower-level obligations often produces lemmas that are inductive relative to several successive frames. If such lemmas were propagated through all levels up to $n$ before the top-level search proceeds, they might already exclude the bad state $b$ or eliminate predecessors that lead to $b$.
In other words, information produced while blocking obligations at lower levels can render some top-level proof obligations obsolete, but in conventional Lemma Sharing this information may arrive too late to prevent redundant work.

\begin{algorithm}[!htbp]
\caption{\textsc{ARPOSBlockController}}\label{alg:ARPOS-block-controller}
\begin{algorithmic}[1]  % [1] = line numbers
\Require trace $(F_0, F_1,\dots, F_n)$, proof obligations queue $\mathcal{Q}$, worker pool $\mathcal{W}$
\While{true}
    \If{$w:=\Call{GetIdleWorker}{\mathcal{W}}$}
        \If{$\langle s, \ell\rangle:=\Call{Peek}{\mathcal{Q}}$ and $\ell \leq n$}
            \State$\Call{Pop}{\mathcal{Q}}$
            \State$\Call{SendMessage}{w,\langle s, \ell\rangle}$
        \Else
            \State$\Call{SendMessage}{w,\mathrm{top\_level}}$\label{ln:send-top-lev-instr}
        \EndIf
    \EndIf
    \State$inbox\gets\Call{ReceiveMessages}{\,}$
    \If{Success $\in inbox$ }
        \State\textbf{return} Success
    \ElsIf{Unsafe $\in inbox$}
        \State\textbf{return} Unsafe
    \EndIf
    \State$\Call{UpdateIdleWorkers}{inbox,\mathcal{W}}$
\EndWhile
\end{algorithmic}
\end{algorithm}

Hence, it might be advantageous to prioritize the use of available parallel resources for blocking already discovered proof obligations at all levels, rather than searching for new top-level proof obligations. To this end, we examine the sequential \textsc{Block} procedure shown in Algorithm \ref{alg:block}. Recall that, whenever a proof obligations is successfully blocked, it is rescheduled at a higher frame (line \ref{ln:block-reschedule} of Algorithm \ref{alg:block}). As a result, multiple proof obligations occupy the same level in the queue $\mathcal{Q}$. Since proof obligations at the same level can be processed in arbitrary order without affecting correctness, they form a natural source of parallelism and can, in principle, be handled concurrently.

Implementing this idea efficiently in a distributed environment requires addressing several challenges. In particular, we must ensure that
\begin{enumerate}[(i)]
    \item each engine makes as much progress as possible given its current local view of the computation state,
    \item work is distributed amongst the engines in a manner that minimizes redundant computation, and
    \item communication remains non-blocking, so that synchronization does not become a bottleneck.
\end{enumerate}
To guide the design of such a scheme, observe that, in the absence of rescheduling, the block procedure \textsc{Block} can be viewed as a recursive depth-first traversal of a tree in which each node corresponds to a proof obligation $\langle s, \ell\rangle $, and whose children are obligations of the form $\langle m, \ell-1\rangle$ added in line \ref{ln:block-child-po} of Algorithm \ref{alg:block}.

\begin{algorithm}[!htbp]
\caption{\textsc{ARPOSBlockWorker}}\label{alg:ARPOS-block-worker}
\begin{algorithmic}[1]  % [1] = line numbers
\Require trace $(F_0, F_1,\dots, F_n)$, property $P$, controller identifier $ctrl$
\While{true}
    \State $inbox\gets\Call{ReceiveMessages}{\,}$
    \If{$\langle s, \ell\rangle\in inbox$}
        \State $po \gets \langle s, \ell\rangle$
    \ElsIf{$\mathrm{top\_level} \in inbox$}
        \If{$\mathsf{SAT}(F_{n} \wedge \neg P)$ with model $m$}\label{ln:worker-top-lev-query}
            \State $po \gets \langle m, n\rangle$
        \Else
            \State$\Call{SendMessage}{ctrl,\mathrm{Success}}$
            \State\textbf{return}
        \EndIf
    \EndIf
    \If{$po$ has a value}
        \State $res \gets \Call{BlockRecursive}{po}$
        \If{$res = \mathrm{Unsafe}$}
            \State$\Call{SendMessage}{ctrl,\mathrm{Unsafe}}$
            \State\textbf{return}
        \Else
            \State$\Call{SendMessage}{ctrl,\mathrm{Idle}}$
        \EndIf
    \EndIf
\EndWhile
\end{algorithmic}
\end{algorithm}

\begin{algorithm}[!htbp]
\caption{\textsc{BlockRecursive}}\label{alg:ARPOS-block-rec}
\begin{algorithmic}[1]  % [1] = line numbers
\Require trace $(F_0, F_1,\dots, F_n)$, proof obligation $\langle s, \ell \rangle$, controller identifier $ctrl$
\State $res \gets \Call{HandleOne}{\langle s,\ell\rangle}$
\If{$res=$ Unsafe}
    \State\textbf{return} Unsafe
%\ElsIf{$res=$ Pushed($i$)}
%\If{$res=$ Pushed($i$)}
%    \State$\Call{Add}{\mathcal{Q}, \langle s, i+1\rangle}$
\ElsIf{$res=$ NewPO$(m)$}\label{ln:block-rec-newpo}
    \If{$\Call{BlockRecursive}{\langle m,\ell- 1\rangle}=$ Unsafe}
        \State\textbf{return} Unsafe
    \Else
        \State\textbf{return} $\Call{BlockRecursive}{\langle s, \ell \rangle}$
    \EndIf
\ElsIf{$res=$ Generalized$(c,k)$}\label{ln:block-rec-generalized}
    \For{$i=1$ to $k$}
        \State $F_i \gets F_i \cup \{ \neg c \}$
    \EndFor
    \State $\Call{SendMessage}{ctrl, \langle s, k+1\rangle}$\label{ln:send-po-to-ctrl}
    \State$\Call{SendLemmaToAll}{c,k}$
\EndIf
\State\textbf{return} Success
\end{algorithmic}
\end{algorithm}

This perspective suggests a natural parallelization strategy. In the parallel scenario, we adopt the recursive formulation of \textsc{Block}, but replace local rescheduling of proof obligations with insertion into a shared priority queue. Proof obligations are then dynamically assigned from this queue to available processors.

We implement this design in our ARPOS strategy using a controller-worker architecture. The controller maintains the shared queue of proof obligations as well as the pool of available workers. Workers retrieve proof obligations, attempt to solve them, and return newly generated obligations to the controller. The pseudocode for the controller is given in Algorithm \ref{alg:ARPOS-block-controller}, while the worker logic is divided into an event-loop component (Algorithm \ref{alg:ARPOS-block-worker}) and the recursive blocking procedure (Algorithm \ref{alg:ARPOS-block-rec}).

In ARPOS, when a new proof obligation is generated at level $\ell-1$ (line \ref{ln:block-rec-newpo} of Algorithm \ref{alg:ARPOS-block-rec}), it is not immediately sent to the controller. Instead, the generating engine first needs to block it by working through the ``local priority queue'' implicit in the recursive formulation. After the proof obligation has been blocked locally, it is sent to the controller, rescheduled at level $\ell$, where it becomes part of the explicit global priority queue of proof obligations that can be scheduled on any available engine.

Recall that our \emph{preemptive propagation} optimization (Section \ref{sec:preempt-prop}) also relies on a controller-worker architecture to ensure well-defined propagation semantics, and therefore integrates naturally with ARPOS.

\subsection{Comparison with other methods}\label{sec:comp-w-other-methods}

Our proposed ARPOS strategy shares similarities with both Lemma Sharing (Section \ref{sec:lemma-sharing}) and PS-PDR \cite{clifton2022fast}, yet differs from each in several important aspects. Like PS-PDR, ARPOS uses a shared queue to manage proof obligations. However, the computational task performed by a worker upon receiving a proof obligation is substantially different.

In PD-PDR, a worker performs a single call to \textsc{HandleOne} and returns the result to the controller. Consequently, if blocking fails, exactly one new proof obligation is generated and reinserted into the queue. In contrast, an ARPOS worker treats a received proof obligation as the root of a recursive blocking process corresponding to an implicit local priority queue. The worker attempts to block all obligations that stem from it, either succeeding completely or returning a counterexample to the overall model-checking problem. Thus, during the execution of a single ARPOS task, multiple new proof obligations may be generated and communicated to the controller (line \ref{ln:send-po-to-ctrl} of Algorithm \ref{alg:ARPOS-block-rec}).

A further important difference concerns the handling of top-level proof obligations. In ARPOS, once the controller exhausts the shared queue, it instructs the next idle worker to search for a new top-level proof obligation (line \ref{ln:send-top-lev-instr} of Algorithm \ref{alg:ARPOS-block-controller} and line \ref{ln:worker-top-lev-query} of Algorithm \ref{alg:ARPOS-block-worker}). In this respect, ARPOS more closely resembles Lemma Sharing, since workers are assigned new tasks immediately upon becoming idle, ensuring continuous utilization of available parallel resources.

\section{Implementation}\label{sec:implementation}

The only work in the literature that systematically studies Lemma Sharing in a distributed setting is \cite{marescotti2017}. In that work, Lemma Sharing is implemented using a client-server architecture in which the engines act as clients that periodically push lemmas to, and pull lemmas from, a central server maintaining a database of all generated lemmas.

In contrast, our implementation of both classical Lemma Sharing and ARPOS (with preemptive propagation) is built on a Rust-based framework for distributed point-to-point communication over MPI, which we integrate with rIC3, a recent high-performance PDR engine written in Rust and the winning entry of HWMCC \cite{hwmcc} in 2025. 
Communication in our framework is asynchronous, relying on non-blocking MPI primitives to overlap communication with computation and thereby improve efficiency. To reduce communication initiation overhead, information can be aggregated into larger messages. Processes probe for incoming messages at user-designated points during the computation. Both the message size and the probing frequency are configurable, allowing communication overhead to be balanced against computation.
The framework is not specific to the strategies covered in this work and can serve more generally as a communication layer for future research on parallel Rust-based PDR engines.

%\red{Future research: It would be interesting to apply the asynchronous rescheduled proof obligation sharing technique proposed here to the likewise PDR-based Quip algorithm \cite{PushingToTheTop2015}. Since both ARPOS and Quip algorithm prioritize trying to push existing lemmas over discovering new bad states, the hope is that the ideas would synergize for a particularly effective parallelization.}

\section{Experiments}\label{sec:experiments}
All experiments were conducted on a cluster consisting of two compute nodes, each equipped with two Intel Xeon Gold 6238T CPUs, 192 GB of RAM, and a 100 Gbps InfiniBand network adapter.
We use the bit-level safety track HWMCC 2025 benchmark suite \cite{hwmcc} in all our experiments and, following standard practice, we impose a wall-clock timeout of 3600 seconds and a per-engine memory limit of 16GB.

The full benchmark suite consists of 319 model-checking instances. We evaluated and compared the following PDR strategies built on the rIC3 engine \cite{rIC3-2025} using our messaging framework as described in Section \ref{sec:implementation}:
\begin{enumerate}
    \item a single rIC3 engine enhanced with CTG \cite{CTG2013} of depth 3,
    \item a portfolio of 16 independent rIC3 engines with CTG, each initialized with a different random seed,
    \item a Lemma Sharing portfolio of rIC3 engines with CTG, consisting of 16 processes on 1 node, 32 processes on 1 node, and 64 processes on 2 nodes, and
    \item an ARPOS portfolio of rIC3 engines with CTG, using the same process configurations as Lemma Sharing.
\end{enumerate}
Due to MPI initialization and communication overhead, ARPOS and Lemma Sharing are at a disadvantage relative to Portfolio on small instances. We therefore focus our comparison on a subset of the benchmark suite, which we denote by \emph{over10}. This subset contains 153 instances for which Portfolio required more than 10 seconds or timed out, but at least one of the evaluated strategies solved the instance within the time limit.

\begin{figure}[ht]
  \centering
%   \begin{subfigure}{0.47\textwidth}
%     \centering
%     \includegraphics[width=\linewidth]{cumulative_time_sharq_big.png}
%     \caption{\emph{more10}}
%   \end{subfigure}
%   \hfill
  \begin{subfigure}{0.5\textwidth}
    \centering
    \includegraphics[width=\linewidth]{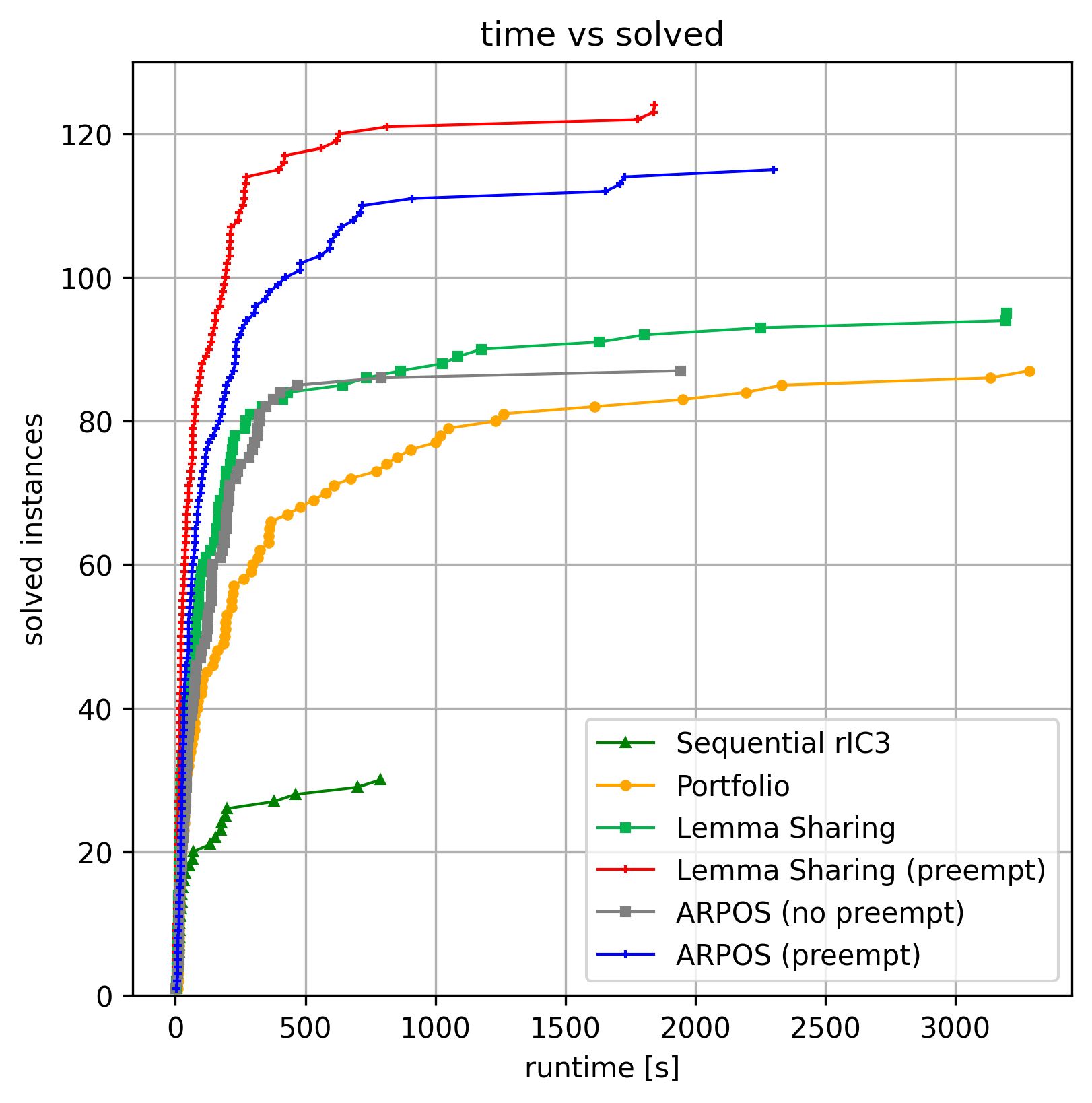}
    % \caption{\emph{less10}}
  \end{subfigure}
  \caption{Number of \emph{over10} instances solved within a given runtime by the strategies Sequential rIC3, Portfolio, as well as Lemma Sharing and ARPOS, both on 16 processes, with and without preemptive propagation.}\label{fig:cactus-16}
\end{figure}

\subsection{Evaluation of preemptive propagation}
In our first series of experiments, we compare the Lemma Sharing and ARPOS strategies, both with and without preemptive propagation, against the standard Portfolio strategy---all using 16 processes on one node on the \emph{over10} data set. The cumulative solved instances for a given runtime cutoff are shown in Figure \ref{fig:cactus-16} and the summary statistics are included in Table \ref{tab:summary-stats-all}. The reported speedup is defined as the ratio between the cumulative runtime of the Portfolio method over the entire data set and the cumulative runtime of the corresponding method.

\begin{figure}[!ht]
  \centering
  \begin{subfigure}{0.47\textwidth}
    \centering
    \includegraphics[width=\linewidth]{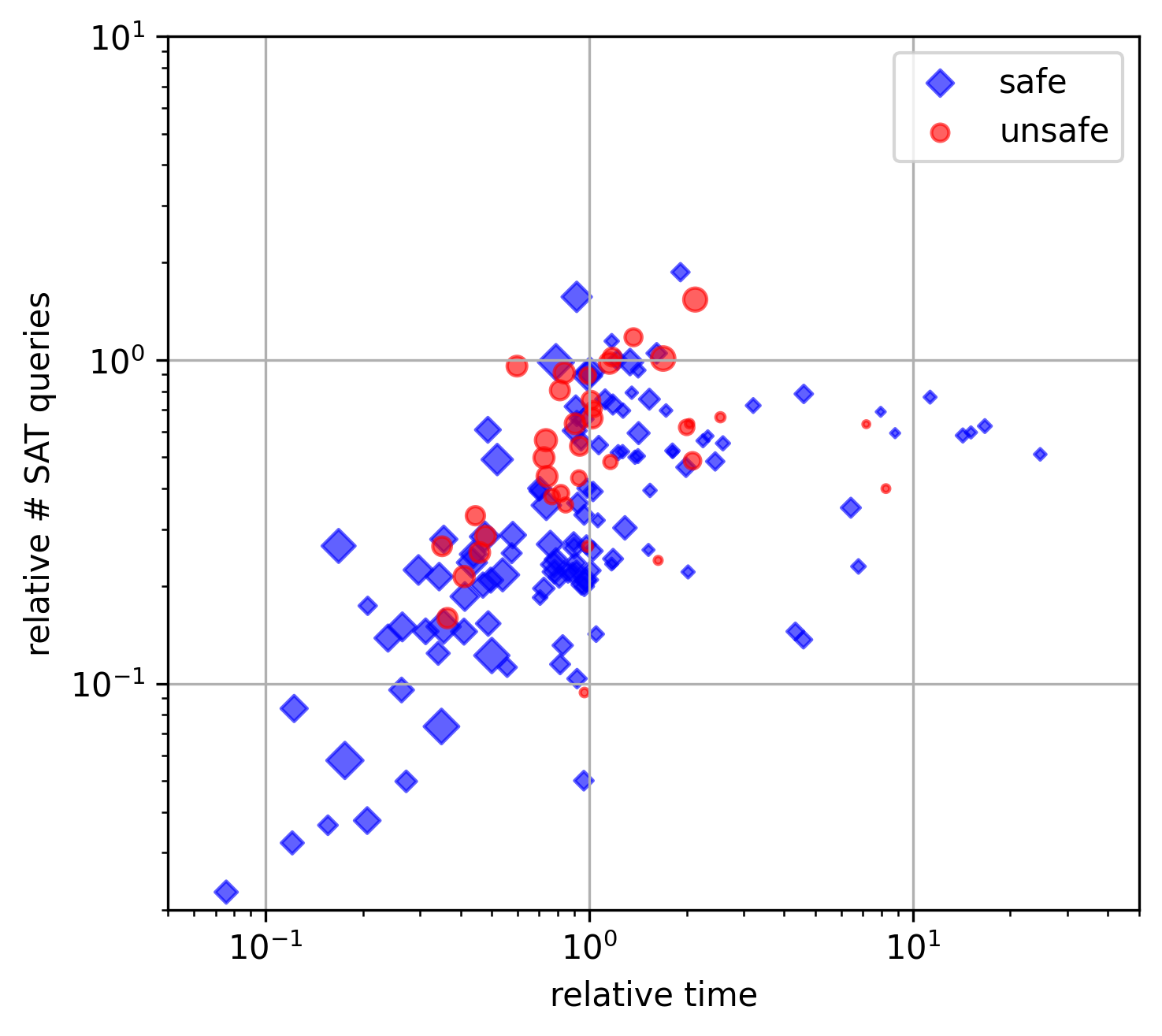}
    \caption{Ratio of Lemma Sharing without preemptive propagation to Lemma Sharing with preemptive propagation.}
  \end{subfigure}
  \hfill
  \begin{subfigure}{0.47\textwidth}
    \centering
    \includegraphics[width=\linewidth]{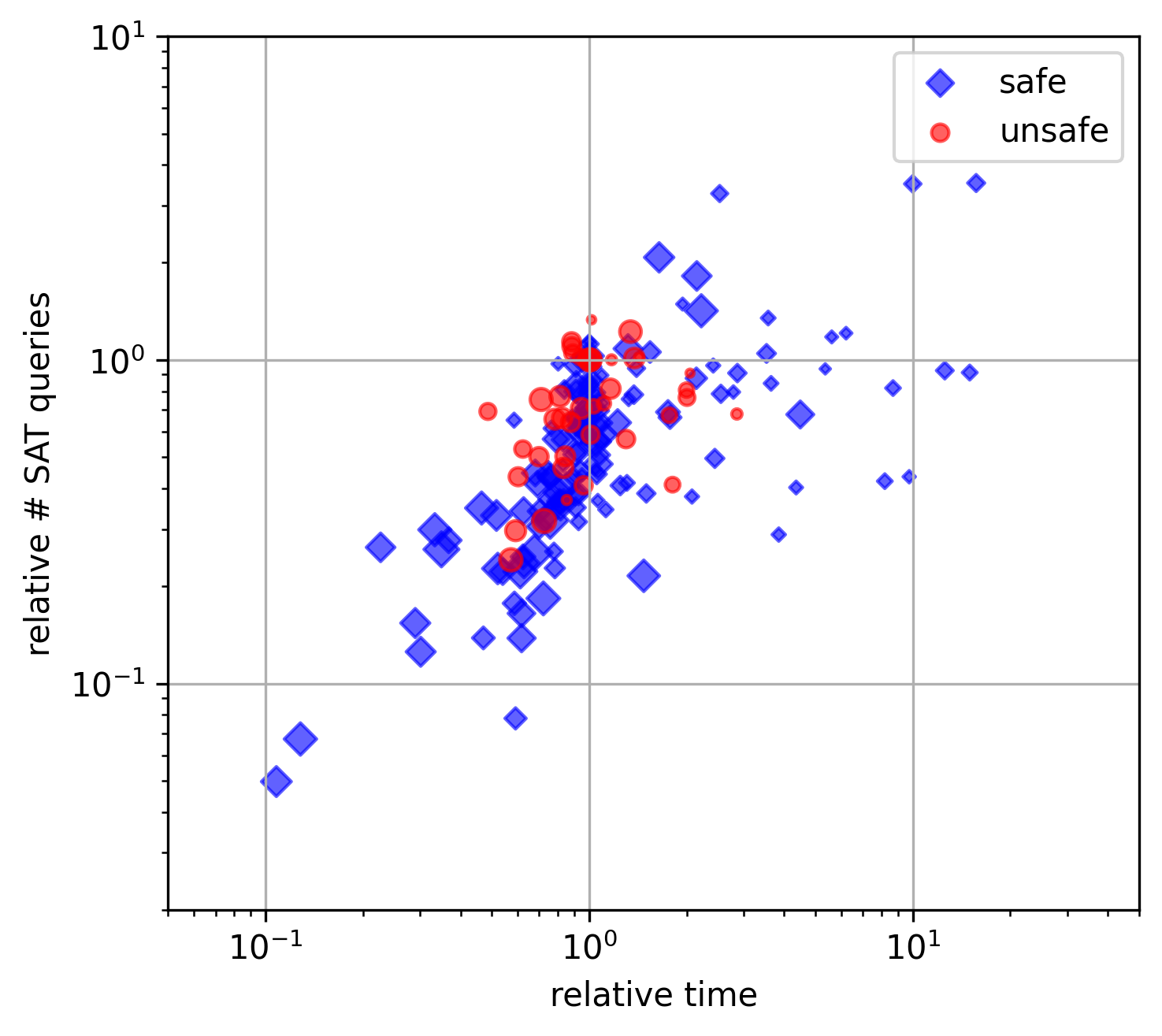}
    \caption{Ratio of ARPOS without preemptive propagation vs ARPOS with preemptive propagation.}
  \end{subfigure}
  \caption{
  Each point represents a model checking instance. For a given PDR strategy, the x-axis shows the ratio of runtime with preemptive propagation to that without it, and the y-axis shows the corresponding ratio of the number of per-engine SAT queries.
  We repeat this for Lemma Sharing and ARPOS. The point size is proportional to the (absolute) runtime.
  }\label{fig:queries-time-each-problem-16}
\end{figure}

Both Lemma Sharing and ARPOS achieve substantial speedups over Portfolio even without preemptive propagation. Enabling preemptive propagation results in an almost $2\times$ additional speedup for Lemma Sharing, however it does not benefit the performance of ARPOS for this processor count.

The speed-up effect of preemptive propagation is further illustrated in Figure~\ref{fig:queries-time-each-problem-16}, which compares both the runtime improvement and the relative change in the number of per-engine issued SAT queries when enabling preemptive propagation for Lemma Sharing and ARPOS. Although runtime measurements for individual PDR instances exhibit the well-known high variability, the experiments demonstrate a clear speed-up trend for the larger benchmarks in the case of Lemma Sharing. Moreover, instances that benefit most from preemptive propagation show a corresponding reduction in the total number of per-engine SAT queries.

% \begin{table*}[ht]
% \centering
% \begin{tabular}{|c|c|c|c|c|c|c|}
% \hline
% Method & Unsafe & Safe & Timed-out & Unique & Best & Avg. speedup \\
% \hline Portfolio & 24 & 63 & 25 & 0 & 16 & \multirow{2}{*}{$2.68\times$}  \\
% Lemma sh. (preempt) & 31 & 68 & 13 & 12 & 83 &\\
% \hline
%  Portfolio & 24 & 63 & 25 & 1 & 18 & \multirow{2}{*}{$2.67\times$}  \\
% ARPOS (preempt) & 30 & 67 & 15 & 11 & 80 &\\
% \hline
% \end{tabular}
% \caption{\red{We should include all methods here.} Summary statistics for \emph{lemma sharing} vs \emph{portfolio} and \emph{ARPOS} vs \emph{portfolio}  strategies on 16 processes on the \emph{over10} data set.}\label{tab:summary-stats-16}
% \end{table*}

\begin{table*}[ht]
\centering
\begin{tabular}{|c|c|c|c|c|c|c|c|}
\hline
Method & \parbox[c][1cm][c]{2cm}{$\#$ nodes,\\$\#$ processes} & \parbox[c][1cm][c]{2cm}{Preemptive propagation} & Unsafe & Safe & Timed-out & Best & Speedup\\
\hline
Portfolio & 1, 16 &  & 24 & 63 & 66 & 4 &   \\
Lemma Sharing & 1, 16 & no & 31 & 64 & 58 & 8 & $1.16\times$  \\
Lemma Sharing & 1, 16 & yes & 35 & 89 & 29 & \textbf{\underline{42}} & $2.23\times$  \\
ARPOS & 1, 16 & no & 34 & 86 & 33 & 10 & $1.83\times$  \\
ARPOS & 1, 16 & yes & 34 & 81 & 38 & 5 & $1.69\times$  \\
\hline
Lemma Sharing & 1, 32 & no & 35 & 85 & 33 & 13 & $1.94\times$  \\
Lemma Sharing & 1, 32 & yes & 35 & 87 & 31 & 31 & $2.12\times$  \\
ARPOS & 1, 32 & no & 33 & 77 & 43 & 5 & $1.47\times$  \\
ARPOS & 1, 32 & yes & 34 & 73 & 46 & 8 & $1.49\times$  \\
\hline
Lemma Sharing & 2, 64 & no & 35 & 89 & 29 & 7 & $2.17\times$  \\
Lemma Sharing & 2, 64 & yes & 35 & 93 & \textbf{\underline{25}} & 19 & \textbf{\underline{2.55$\times$}}  \\
ARPOS & 2, 64 & no & 31 & 79 & 44  & 11 & $1.56\times$  \\
ARPOS & 2, 64 & yes & 34 & 75 & 45  & 10 & $1.55\times$  \\
\hline
\end{tabular}
\caption{Summary statistics comparing Lemma Sharing and ARPOS against Portfolio on the \emph{over10} data set. Speedup is defined as the ratio between the cumulative runtime of the Portfolio method on the whole data set and the corresponding runtime of the given method.}\label{tab:summary-stats-all}
\end{table*}

\subsection{Scaling to large process counts}
\begin{figure}[ht]
  \centering
%   \begin{subfigure}{0.47\textwidth}
%     \centering
%     \includegraphics[width=\linewidth]{cumulative_time_sharq_big.png}
%     \caption{\emph{more10}}
%   \end{subfigure}
%   \hfill
  \begin{subfigure}{0.5\textwidth}
    \centering
    \includegraphics[width=\linewidth]{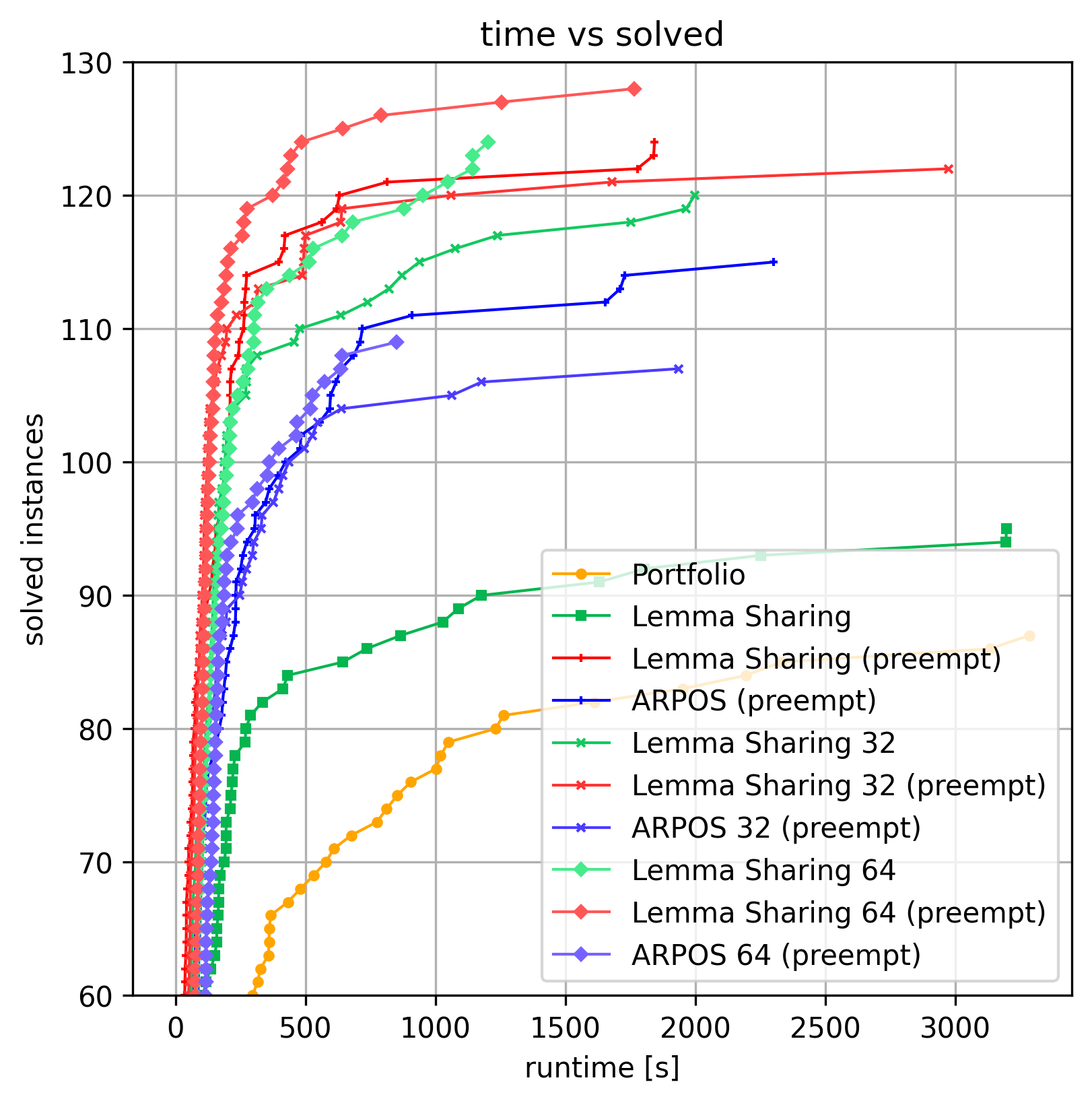}
    % \caption{\emph{less10}}
  \end{subfigure}
  \caption{Runtime comparison of Lemma Sharing and ARPOS, each on 16, 32, and 64 processes.}\label{fig:scaling-16-64}
\end{figure}

Next, we compare the performance of Lemma Sharing and ARPOS with preemptive propagation enabled using larger numbers of engines in a distributed setting. The results of these experiments are summarized in Table \ref{tab:summary-stats-all}, and the cumulative plots are shown in Figure
\ref{fig:scaling-16-64}.

The Lemma Sharing strategies yield a substantive speedup over the Portfolio for all process counts, and in each case enabling preemptive propagation significantly boosts the performance even further, with the largest boost observed for 16 processes where the speed-up jumps from $1.16\times$ to $2.23\times$. This configuration was also best-performing for the largest number of benchmarks. We recorded the overall largest speed-up of $2.55\times$ for the Lemma Sharing strategy on 64 processes with preemptive propagation enabled.

We observe that the ARPOS strategy does not yield speed-ups competitive with Lemma Sharing, even though it was the best-performing strategy for a number of benchmarks. 
At this point we do not have a convincing explanation for this performance gap, highlighting the challenge of gaining a deeper understanding of the scalability characteristics of parallel PDR methods in future work. 

% \begin{figure}[ht]
%   \centering
%   \begin{subfigure}{0.47\textwidth}
%     \centering
%     \includegraphics[width=\linewidth]{speedup_scatter_Portfolio_Lemma Sharing (preempt).png}
%     \caption{\emph{lemma sharing} with preemptive propagation vs. \emph{portfolio}}
%   \end{subfigure}
%   \hfill
%   \begin{subfigure}{0.47\textwidth}
%     \centering
%     \includegraphics[width=\linewidth]{speedup_scatter_Portfolio_ARPOS (preempt).png}
%     \caption{\emph{ARPOS} with preemptive propagation vs. \emph{portfolio}}
%   \end{subfigure}
%   \caption{Runtime comparison for \emph{portfolio}, \emph{lemma sharing}, and \emph{ARPOS}, each on 16 processes.}\label{fig:runtime-each-problem-16}
% \end{figure}

\section{Conclusion and Future Work}\label{sec:conclusion}
We have presented ARPOS, a novel sharing-based parallelization strategy for the PDR algorithm that improves scalability to large processor counts compared to classical Lemma Sharing on hardware model checking problems. The key idea behind ARPOS is to maintain a global queue of rescheduled proof obligations, enabling parallel resources to be directed toward resolving rescheduled proof obligations rather than searching for new top-level proof obligations.
we implemented ARPOS on top of the state-of-the-art rIC3 hardware model checker using an MPI-based asynchronous message passing framework.

Scaling model checking algorithms to meet the demands of increasingly complex industrial verification problems remains a significant theoretical and engineering challenge. We hope that the ideas introduced in this work will help advance the understanding of scalable parallelization techniques for PDR and inspire further research into parallel and distributed approaches to model checking more broadly. An immediate next step would be to explore the effect of message size when aggregating lemmas and proof obligations to reduce communication initiation overhead, both for Lemma Sharing and ARPOS. Moreover, ARPOS could be evaluated on benchmarks other than HWMCC to investigate if the problem type affects the relative performance of different methods.

\subsubsection*{Acnknowledgments}\label{sec:acknowledgements}
I am grateful to Kiril Dichev, Petro Lutsyk, Bill McColl, and Jonas Oberhauser for helpful discussions. 

\printbibliography

@InProceedings{kinduction2000,
author="Sheeran, Mary
and Singh, Satnam
and St{\aa}lmarck, Gunnar",
editor="Hunt, Warren A.
and Johnson, Steven D.",
title="Checking Safety Properties Using Induction and a SAT-Solver",
booktitle="Formal Methods in Computer-Aided Design",
year="2000",
publisher="Springer Berlin Heidelberg",
address="Berlin, Heidelberg",
pages="127--144",
isbn="978-3-540-40922-9"
}

@online{hwmcc,
  author    = {HWMCC},
  url       = {https://hwmcc.github.io/}
}

@inproceedings{Bradley2011SATBasedMC,
  title={SAT-Based Model Checking without Unrolling},
  author={Aaron R. Bradley},
  booktitle={International Conference on Verification, Model Checking and Abstract Interpretation},
  year={2011},
  url={https://api.semanticscholar.org/CorpusID:17018335}
}

@INPROCEEDINGS{EenPDR2011,
  author={Een, Niklas and Mishchenko, Alan and Brayton, Robert},
  booktitle={2011 Formal Methods in Computer-Aided Design (FMCAD)}, 
  title={Efficient implementation of property directed reachability}, 
  year={2011},
  volume={},
  number={},
  pages={125-134},
  doi={}}

@inproceedings{PushingToTheTop2015,
author = {Ivrii, Alexander and Gurfinkel, Arie},
title = {Pushing to the top},
year = {2015},
isbn = {9780983567851},
publisher = {FMCAD Inc},
address = {Austin, Texas},
booktitle = {Proceedings of the 15th Conference on Formal Methods in Computer-Aided Design},
pages = {65–72},
numpages = {8},
location = {Austin, Texas},
series = {FMCAD '15}
}

@INPROCEEDINGS{CTG2013,
  author={Hassan, Zyad and Bradley, Aaron R. and Somenzi, Fabio},
  booktitle={2013 Formal Methods in Computer-Aided Design}, 
  title={Better generalization in IC3}, 
  year={2013},
  volume={},
  number={},
  pages={157-164},
  doi={10.1109/FMCAD.2013.6679405}}

@INPROCEEDINGS{InternalSignals2021,
  author={Dureja, Rohit and Gurfinkel, Arie and Ivrii, Alexander and Vizel, Yakir},
  booktitle={2021 Formal Methods in Computer Aided Design (FMCAD)}, 
  title={IC3 with Internal Signals}, 
  year={2021},
  volume={},
  number={},
  pages={63-71},
  doi={10.34727/2021/isbn.978-3-85448-046-4_14}}

@inproceedings{chakikarimi2016,
author = {Chaki, Sagar and Karimi, Derrick},
title = {Model Checking with Multi-threaded IC3 Portfolios},
year = {2016},
isbn = {9783662491218},
publisher = {Springer-Verlag},
address = {Berlin, Heidelberg},
url = {https://doi.org/10.1007/978-3-662-49122-5_25},
doi = {10.1007/978-3-662-49122-5_25},
booktitle = {Proceedings of the 17th International Conference on Verification, Model Checking, and Abstract Interpretation - Volume 9583},
pages = {517–535},
numpages = {19},
location = {St. Petersburg, FL, USA},
series = {VMCAI 2016}
}

@InProceedings{marescotti2017,
  author={Marescotti, Matteo and Gurfinkel, Arie and Hyvärinen, Antti E. J. and Sharygina, Natasha},
  booktitle={2017 Formal Methods in Computer Aided Design (FMCAD)}, 
  title={Designing parallel PDR}, 
  year={2017},
  volume={},
  number={},
  pages={156-163},
  doi={10.23919/FMCAD.2017.8102254}}

@inproceedings{rIC3-2025,
author = {Su, Yuheng and Yang, Qiusong and Ci, Yiwei and Bu, Tianjun and Huang, Ziyu},
title = {The rIC3 Hardware Model Checker},
year = {2025},
isbn = {978-3-031-98667-3},
publisher = {Springer-Verlag},
address = {Berlin, Heidelberg},
url = {https://doi.org/10.1007/978-3-031-98668-0_9},
doi = {10.1007/978-3-031-98668-0_9},
booktitle = {Computer Aided Verification: 37th International Conference, CAV 2025, Zagreb, Croatia, July 23-25, 2025, Proceedings, Part I},
pages = {185–199},
numpages = {15},
location = {Zagreb, Croatia}
}

@inproceedings{LocalizationAbs2017,
  author       = {Yen-Sheng Ho and Alan Mishchenko and Robert Brayton and Niklas E{\r{e}}n},
  title        = {{Enhancing PDR/IC3 with Localization Abstraction}},
  booktitle    = {Proceedings of the International Workshop on Logic and Synthesis (IWLS)},
  year         = {2017},
  address      = {USA},
  note         = {Presented at IWLS 2017},
  url          = {https://people.eecs.berkeley.edu/~alanmi/publications/2017/iwls17_pdr.pdf}
}

@inproceedings{clifton2022fast,
  author    = {Marshall Clifton and Charles Gretton},
  title     = {{Fast Parallel PDR Algorithms for Planning}},
  booktitle = {Proceedings of the ICAPS 2022 Workshop on Knowledge Engineering for Planning and Scheduling (KEPS)},
  year      = {2022},
  address   = {Virtual / Canberra, Australia},
  note      = {Marshall Clifton and Charles Gretton},
}

@article{pdr-auto-planning-2014,
author = {Suda, Martin},
title = {Property directed reachability for automated planning},
year = {2014},
issue_date = {May 2014},
publisher = {AI Access Foundation},
address = {El Segundo, CA, USA},
volume = {50},
number = {1},
issn = {1076-9757},
journal = {J. Artif. Int. Res.},
month = may,
pages = {265–319},
numpages = {55}
}

@InProceedings{pdr-software-2020,
author="Beyer, Dirk
and Dangl, Matthias",
editor="Biere, Armin
and Parker, David",
title="Software Verification with PDR: An Implementation of the State of the Art",
booktitle="Tools and Algorithms for the Construction and Analysis of Systems",
year="2020",
publisher="Springer International Publishing",
address="Cham",
pages="3--21",
isbn="978-3-030-45190-5"
}

@article{komuravelli2016anvesh,
author = {Komuravelli, Anvesh and Gurfinkel, Arie and Chaki, Sagar},
title = {SMT-based model checking for recursive programs},
year = {2016},
issue_date = {June      2016},
publisher = {Kluwer Academic Publishers},
address = {USA},
volume = {48},
number = {3},
issn = {0925-9856},
url = {https://doi.org/10.1007/s10703-016-0249-4},
doi = {10.1007/s10703-016-0249-4},
journal = {Form. Methods Syst. Des.},
month = jun,
pages = {175–205},
numpages = {31}
}

\end{document}